# Design of flexible ultrahigh-$Q$ microcavities in diamond-based photonic crystal slabs

Snjezana Tomljenovic-Hanic[1]*, Andrew D. Greentree[2], C. Martijn de Sterke[1], and Steven Prawer[2]

[1]ARC Centre of Excellence for Ultrahigh-bandwidth Devices for Optical Systems (CUDOS), and School of Physics, University of Sydney, Camperdown, NSW 2006, Australia
[2]School of Physics, University of Melbourne, Melbourne, Victoria 2010, Australia
* Corresponding author: snjezana@physics.usyd.edu.au

**Abstract**: We design extremely flexible ultrahigh-$Q$ diamond-based double-heterostructure photonic crystal slab cavities by modifying the refractive index of the diamond. The refractive index changes needed for ultrahigh-$Q$ cavities with $Q \sim 10^7$, are well within what can be achieved ($\Delta n \sim 0.02$). The cavity modes have relatively small volumes $V < 2\ (\lambda/n)^3$, making them ideal for cavity quantum electro-dynamic applications. Importantly for realistic fabrication, our design is flexible because the range of parameters, cavity length and the index changes, that enables an ultrahigh-$Q$ is quite broad. Furthermore as the index modification is post-processed, an efficient technique to generate cavities around defect centres is achievable, improving prospects for defect-tolerant quantum architectures.

## 1. Introduction

One of the most exciting challenges in quantum optics is the design and fabrication of solid-state devices that generate strong, lossless single-atom single-photon coupling. This challenge will only be met by the combination of photonic engineering techniques with quantum physics applications. To observe the dynamics of strongly coupled atom-photon systems, and to harness these dynamics for practical technologies requires high quality factors, $Q$, and small modal volume, $V$. Since the dynamics of these systems scales as $Q/V^{1/2}$ [1], a small, extremely low loss cavity is needed to reach the strong coupling regime. Photonic crystals appear to be good candidates to control the rate of spontaneous emission since they can suppress any propagation in the structure if the frequency falls within a complete bandgap [2].

Nearly all of the recent photonic crystal cavity studies considered a photonic crystal slab (PCS) composed of silicon or other high-index semiconductors with the refractive index $n\sim3.4$. Such high refractive index is desirable for strong mode confinement, however there are no known convenient large-dipole moment optical centres in the high-index media. In this work we consider microcavities in a PCS made of single crystal diamond. Diamond is unique in the solid state as the host for a wide range of large-dipole, photo-stable and room-temperature active colour centres, for example the well-known negatively charged nitrogen-vacancy (NV) centre, and the nickel related NE8 centre (for a recent review, see [3]). As diamond has a substantially lower refractive index ($n=2.4$), until recently it was unclear if sufficiently high quality factors can be obtained within small mode volume diamond PCS cavities [4]. The importance of combining diamond colour centres with optical cavities has been explored over the last few years, and many approaches to enhancing photon collection from colour centres and atom-photon coupling have been discussed: including nano-crystal diamond on fibre [5,6], nano-diamond on silica microspheres [7], diamond toroidal [8] and photonic crystal microcavities [9]. Recently, a high-$Q$ cavity design using a silicon nitride-based one dimensional photonic crystal cavity for coupling to a diamond nanocrystal was also proposed [10].

Theoretical investigations of diamond photonic bandgap cavities have followed the designs developed in silicon based devices. Preliminary theoretical designs can be found in [4,11,12], and some diamond photonic crystal structures were fabricated [9]. As most of the structures incorporated membrane designs, the need for thin diamond membranes has pushed fabrication limits. Approaches taken to generate membranes of the required thickness include the growth of thin layers of nanocrystalline diamond (as used in [9,13]), and modifications of the lift-off technique using single [14] and double [15] implantation processes.

At present, the geometry of ultrahigh-$Q$ PCS cavities is finalized at the stage of fabrication, requiring changes in the geometry, holes' size and position, with nanometre precision [16,17]. However there are other ways to localize light within a PCS without any geometry perturbation [18-20]. It is possible to form double-heterostructure type cavities by changing the refractive index of the background material [18] or holes [19]. The main advantage of this type of cavity is the possibility of post-processing the cavity around the region of interest. Unlike conventional heterostructures, these designs do not require changes in the geometry with nanometre precision. Here we consider a realistic and extremely flexible double-heterostructure design that was previously applied to photosensitive material-based PCS such as chalcogenide glasses [18]. We design the diamond-based double-heterostructure by changing the refractive index of a slab, which may be achieved by ion-beam irradiation. There are remarkably few studies of the effect of ion bombardment on the refractive index of diamond. Hines [21] shows a monotonic increase in the refractive index of diamond with damage, reporting changes from $n=2.42$ to 2.62 at doses of $20 \times 10^{14}$ ions/cm$^2$ from 20 keV C$^+$ bombardment. On the other hand Bhatia *et al.* [22] reported a *reduction* in the refractive index for low doses, to around $n=2$ for a dose of $1.5 \times 10^{14}$ Carbon ions/cm$^2$. This latter result is particularly important for our purposes as it enables the cavity to be defined in an *undamaged* region, which is expected to improve device properties. As shown below, these refractive index changes are larger than required for our designs. We note that Bhatia's result appears to reflect the low-dose region more accurately and preliminary results at the University of Melbourne seem to support these findings [23].

## 2. The concept

Diamond is a unique material, with a host of attributes that seem to favour it as a platform for solid-state optical approaches to quantum information processing. Amongst the many outstanding properties of diamond, the most important for our applications are that it possesses the largest transparency window in the visible regime, has the highest thermal conductivity, and, most importantly, host a large number of high dipole moment colour centres. At present, the most important of these centres is the negatively charged NV centre as it shows efficient spin polarization [24], room temperature single spin readout

[25], long ground state coherence (350 $\mu$s) [26], and efficient single photon generation [27]. At liquid helium temperatures there have also been demonstrations of transform limited single photons [28] and coherent population trapping [29]. NV centres are relatively easy to create by implanting $N^+$ or $N_2^+$ ions into ultra-low N diamond [30], with good precision. The inherently stochastic nature of ion implantation means that the ultimate *a priori* precision is determined by the straggle of the implanted ions, which depends on implant species and energy. So although it is in principle possible to register each ion strike and to locate the implanted ion to high precision, it is not possible to implant an ion to a location with better than around 20 nm precision for the cases of most interest.

In addition to the NV centre, there are also other centres with promising features, such as the nickel-related NE8 centre [31], the silicon-vacancy centre [32], and xenon related centres [33]. Here we concentrate on the NV centre and cavities resonant with the zero-phonon line transition at 637 nm. However our strategies can be easily applied to the other centres.

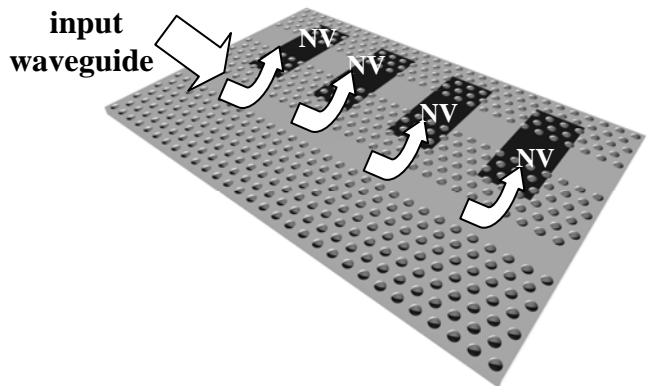

Fig. 1 Concept for heterostructure cavity array coupled to a common control line. Line defects acting as waveguides allow for propagating control pulses, with individual atom-cavity systems defined along another line defect. As the NV fabrication is probabilistic, it is preferable to induce the cavities only in the vicinity of post-selected NV centres which are optically active and favorably aligned with respect to the required cavity mode. The dark regions represent cavities.

Despite the remarkable properties of the NV centre, there are issues with its use in scalable quantum architectures. First, the structure is still not completely understood, see [34,35,36] for some of the latest models. More critically, there are pronounced phonon sidebands which greatly limit the possibility for deterministic generation of identical photons from subsequent pulses or different centres. However, it has been shown that the effect of the phonon sidebands can be effectively eliminated by using high-$Q$ cavities with $Q$s much less than are demonstrated here [36]. Finally, there are no known NV production techniques that guarantee unity probability of formation, and even if such a technique were available, there are still four distinct orientations available to the centre, so if a particular orientation is required, a theoretical maximum yield of one in four is the highest that could be achieved. Hence it is clear that any architecture must be able to cope with probabilistic generation of atom-cavity systems, i.e. it must incorporate defect-tolerant manufacturing approaches. One convenient approach is to perform semi-random implantation and post-select centres which have formed with one centre sufficiently isolated from nearby NV centres and random fluctuators, and with the required orientation. Such analyses can in principle be performed post-implantation with relative ease for NV centres, and then create cavities around them using the designs proposed here. This removes the issue of atom-cavity alignment and fabrication variability, and our designs are ideally suited to such post-processing methodologies.

In Fig. 1 we show a concept design for how an array of cavities coupled to a common input waveguide might be realised. The common bus could be used for distributing photons and effecting nonlocal cavity-cavity coupling as shown. The cavities themselves are heterostructure cavities formed about a common line defect, and the design would be optimised to place the cavities only around NV centres which were appropriately placed within the channel and with correct orientation. This approach naturally satisfies the defect tolerant design methodology discussed above.

## 3. Cavity design and method

We design the heterostructure by changing the refractive of the diamond-based PCS. The resulting refractive index is highest in the central part of PCS, indicated by the dark gray region in Fig. 2. Consequently, the central part of the PCS acts as a cavity. First we consider a uniform and step-like refractive index change being introduced across the waveguide direction, as shown in Fig. 2(a). Later we consider a gradual profile in the waveguide direction, shown in Fig. 2(b). This gradual refractive index profile is still step-like but with smaller steps, e.g. smaller refractive index differences at the interfaces. Experimentally the steps can be made either by controlled rastering of the ion-beam source with varying dwell times per region to create the damage profile, or by multiple uniform implantation steps with changes in surface masking between implantation steps to achieve the desired profile. The latter method suffers from alignment issues, although the robustness of the cavity design should ameliorate some of these issues.

Our numerical model is a PCS composed of a hexagonal array of cylindrical air holes in a diamond slab with $n=2.4$. The lattice constant is $a$, hole radius $R=0.29a$ and slab thickness is $h=0.9a$. Across the PCS there is a line defect, W1 waveguide, in the Γ-K direction. A cavity is formed by the inducing the refractive index change outside the central part of the slab that can have a step-index profile shown in Fig. 2(a) or a gradual-index profile shown in Fig. 2(b). As low damage decreases the refractive index, the central part need not be exposed to the radiation and therefore has refractive index, $n=2.4$ whereas outside the cavity the radiation damage reduces the refractive index, $n<2.4$. Here we concentrate on the refractive index change required to provide photonic confinement. More importantly, one might expect residual damage to be a source of decoherence for the atomic system, and so by maintaining a pristine environment around the centre we ensure the best possible quantum properties for the system.

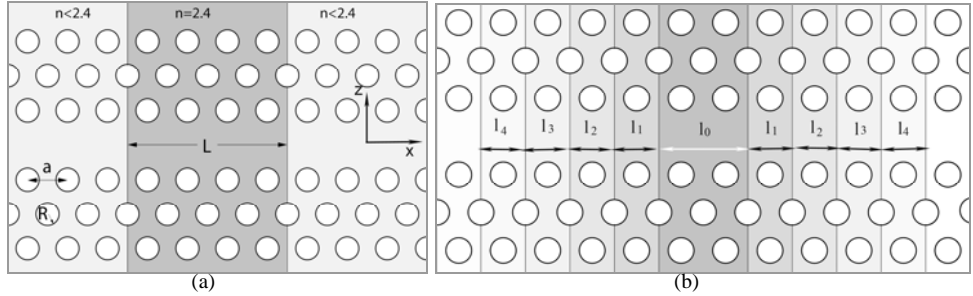

Fig. 2 Schematic of the PCS considered here with (a) a step-index profile and (b) a gradual-index profile. The undamaged region in the centre has the largest refractive index, whilst damage reduces the refractive index in the surrounding regions, defining the cavity. The cavity length is multiple of periods $L=ma$, where $m$ is an integer for the step-profile and $L=l_0+2(l_1+l_2+l_3+l_4)$, where $l_i=ma$ for the gradual-profile.

The PCS structure has 33 to 35 periods in the $x$-direction and 25 periods in the $z$-direction for the step profile structure. The gradual profile structure has 37 periods in the $x$-direction and 27 in the $z$-direction because the cavity is longer. The size of the structure is dictated by the size of the cavity. Special care has to be taken when choosing a size of the entire structure. In particular the dimension in the waveguide direction has to be chosen carefully as otherwise light may be lost via the waveguide, thus lowering the cavity's $Q$. If the PCS is not long enough in comparison to the cavity length, the in-plane component can be a limiting factor [20]. In all cases considered here the size of the structure was dictated by a condition that in-plane losses are not a limiting factor. The cavity size is defined by, $L=ma$, where $m$ is an integer for the step-index profile and $L=l_0+2(l_1+l_2+l_3+l_4)$ for the gradual-index profile where $l_i=ma$. We assume that the induced refractive index change is homogenous within the slab in the vertical direction, and hence this implies damage by light ion (e.g. proton) bombardment.

The optimal positioning of the NV centre would be at the maximum of the cavity mode. This ensures that centre-photon coupling will be largest. In practise there are experimental limitations to how well the cavity mode and centre can be aligned. Such limitations arise from the fundamental straggle range of the implanted ions and the alignment of the damage post-processing to create the cavity. By using implantation techniques similar to those developed for counted single phosphorus ion implantation in silicon [35] it should be possible to implant nitrogen ions to a precision around 20 nm, which will, with high probability, place the NV centre near a maximum of the cavity electric field. Such variability ultimately requires that all atom-cavity systems produced be characterised to determine the actual coupling in the fabrication of a large-scale structure. Note that most of the resonant mode field is concentrated within a much larger region with a typical size of a few 100s of nm.

We analyse the band structure and mode-gap of the structure using the plane wave expansion method (PWEM). The quality factor, $Q$, is calculated using finite-difference time-domain (FDTD) simulation. Computational parameters are similar to those used in Ref [19]. For the step-index profile the optimal refractive index change for a given cavity length is obtained considering the relative position of the resonant mode within the mode-gap [19]. The gradual-index profile cavities are not fully optimized.

## 4. Numerical results

The design of high-$Q$ cavities based on double-heterostructures relies on the mode-gap effect that provides light localization within a part of the waveguide [17-20]. Therefore first we examine the dispersion curves of the homogenous structure using the PWE method. The results for dispersion curves for slabs with $n=2.4$ and $n=2.38$, are shown in Fig. 3(a). Within the lowest photonic band gap (PBG) there are two modes of even-like symmetry. The lower mode is of interest as it provides higher-$Q$ cavities. By decreasing the refractive index of the slab, the dispersion curves move up towards higher frequencies. The mode-gap is defined by the range of frequencies between those two dispersion curves. If the structures with two refractive indices are combined in a double-heterostructure, there is possibility to localize light within the central, non-damaged region, due to the mode-gap effect.

We combine those PCSs in a single structure as shown in Fig. 2(a) and calculate quality factors as a function of the induced refractive index change for fixed cavity length. In Fig. 3(b) the quality factor is plotted as a function of the refractive index change that ranges from $\Delta n=0.01$ to $\Delta n=0.03$. This refractive index range was chosen according to the analysis in Ref [19], and is compatible with the lowering of the refractive index reported in the low-dose region Bhatia et al [22], which exceeded these amounts by around one order of magnitude. The cavity length is fixed at $L=4a$ and $L=6a$.

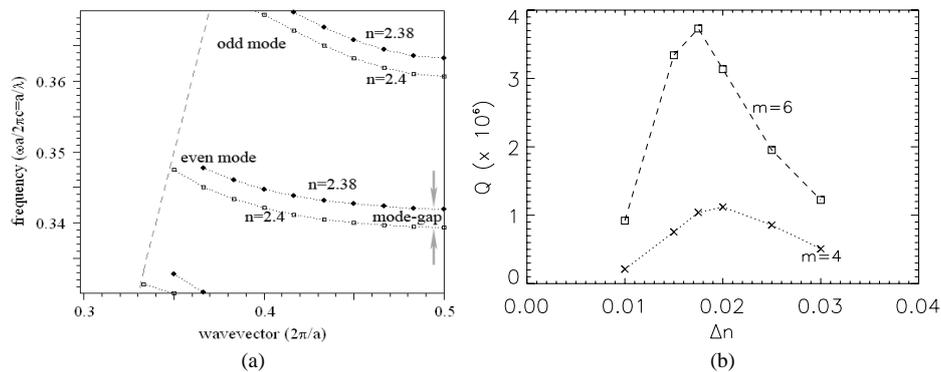

Fig. 3 (a) Dispersion curves for the regular (empty rectangles) and a structure with damaged induced refractive index change (full circles); (b) total quality factor as a function of the refractive index change for the step profile for $m=4$ (crosses) and $m=6$ (rectangles).

At the refractive index difference of $\Delta n=0.01$ the quality factors are $Q=2.1\times10^5$ for $m=4$ and $Q=9.2\times10^5$ for $m=6$. As expected a larger cavity has a larger $Q$ for the same refractive index change. For both cavities, $Q$ increases as the refractive index difference is increased. For the shorter cavity the maximum value of $Q=1.1\times10^6$ appears at $\Delta n=0.02$. A further increase in the refractive index decreases the quality factor but it still remains at the order of few $10^5$. The fact that the optimum occurs at $\Delta n=0.02$ agrees with our previous findings for other heterostructure designs [19]. Similar findings apply for the longer cavity, $m=6$. The optimal quality factor, $Q=3.7\times10^6$, appears at $\Delta n=0.0175$. Note that over almost all refractive index range, the quality factor is $Q\sim10^6$. Comparing the results for those two cavities we can conclude that for longer cavities, smaller refractive index changes are needed to obtain optimal $Q$. However at the same time the modal volume increases. The ratio $Q/V^{1/2}$ is an important factor for quantum applications. Therefore we calculate the modal volumes of the resonant modes at the maximum for both cavities. For the shorter cavity the modal volume is $V=1.73\ (\lambda/n)^3$ and for $m=6$ cavity it is $V=1.94\ (\lambda/n)^3$. The ratio $Q/V^{1/2}$ is still three times larger for the longer cavity. Even though the resonant mode is slightly enlarged the overall ratio is significantly larger due to the larger $Q$.

According to the results presented in Fig. 3(b), this cavity is quite flexible in terms of the refractive index change. This is a significant advantage of these schemes as precise control of the total applied dose can be challenging. Now we investigate sensitivity to variations in the cavity length to quantify the robustness of the design to fabrication imperfections. Therefore we fix the refractive index at $\Delta n=0.0175$ and vary the cavity length around $L=6a$. The results for the quality factor and modal volume are shown in Fig. 4(a). Choosing a periodicity of $a=240$ nm which is appropriate for diamond-based PCS working at the zero-phonon line of NV centres [9], the variation of 100 nm considered here is 100 times larger than precision required for the other designs [4,11,12]. Within this range the quality factor varies between $Q=3.5\times10^6$ for $\Delta L=-100$ nm and $Q=2.7\times10^6$ at $\Delta L=+100$ nm. The quality factor even increases for a cavity that is 50 nm shorter. The variation in modal volume is even much smaller than the quality factor variation. It varies from $V=1.90\ (\lambda/n)^3$ for $\Delta L=-100$ nm to $V=1.99\ (\lambda/n)^3$ at $\Delta L=+100$ nm.

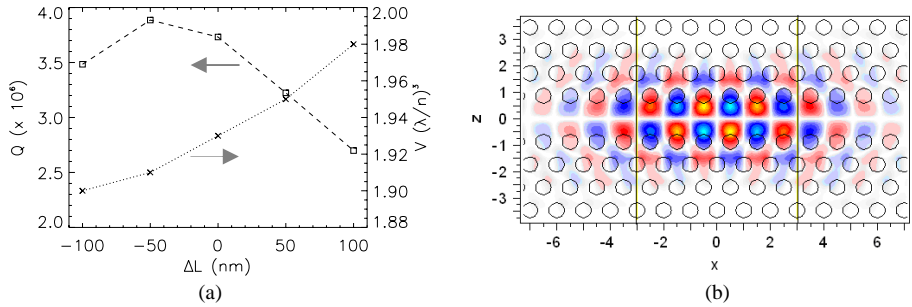

Fig. 4 (a) Total $Q$ (squares) and modal volumes (crosses) for $m=6$ and $\Delta n=0.0175$ cavity as a function of the cavity length variation, the PCS periodicity is $a=240$ nm; (b) the major electric field component, $E_x$, in the middle of the slab. The circles indicate the air-holes and the vertical lines indicate the cavity edges.

Therefore the range of parameters, cavity length and the refractive index change that enable a high-$Q$ cavity is quite broad. Consequently, it is not crucial to control the size and the refractive index change precisely, a vital considerations in the development of practical and robust quantum photonic devices considering the less advanced fabrication tools available to diamond, as compared with silicon processing. The tolerance of our design contrasts with the designs based on geometry perturbation where the high-$Q$ cavity is extremely sensitive to changes of the optimal parameters [4,16,17]. Such robustness aligns well with the practicalities of implanting individual dopants into diamond cavity structures, as mentioned above, and is a strong motivation for experimentally investigating these cavity designs.

To obtain a strong field-matter interaction, the electric field should be mainly concentrated in the diamond slab and not in the air-holes, as the coupling to the diamond colour centres is the paramount aim. In Fig. 4(b) we plot the major electric field component, $E_x$, in the middle of the slab. The holes are indicated by black circles and the cavity edges are indicated by vertical lines. Indeed the field is mainly concentrated in the dielectric. This is expected as the mode-gap is located close to the lower band gap edge, dielectric band [2]. Note that the most of the field is located in non-damaged central part as predicted. For the field plotted in Fig. 4(b) 15 % of the field energy is located in the damaged region. This number can be reduced narrowing the waveguide width [37]. This is important as absorption induced by damaged would otherwise affect the cavity [20].

We finally consider more complicated refractive index profiles. It was recently demonstrated that a gradual rather than a step profile refractive index change increases the quality factor [17,18]. In Ref [17] an intermediate layer between the perturbed PCS and the regular PCS was used. This gradual refractive index profile is still step-like but with smaller steps, e.g. smaller index differences at the interfaces. Cavity fabrication via multiple mask processes is experimentally demanding, but is not beyond the limitations of existing fabrication techniques as the underlying photonic crystal structure can be used as permanent registration markers for subsequent alignment steps.

Table 1. The total quality factor for the gradual index profile

| $\Delta n$ | $l_0 [a]$ | $l_1 [a]$ | $l_2 [a]$ | $l_3 [a]$ | $l_4 [a]$ | $Q$ |
|---|---|---|---|---|---|---|
| 0.0035 | 4 | 1 | 1 | 1 | 1 | $8.4 \times 10^6$ |
| 0.004 | 4 | 2 | 2 | 2 | 2 | $1.8 \times 10^7$ |
| 0.004 | 4 | 1 | 1 | 1 | 1 | $3.1 \times 10^7$ |

Here we briefly consider the gradual profile shown in Fig. 2(b). The central undamaged PCS, with the refractive index $n=2.4$ is $l_0=4a$ wide. The damaged region has steps with the refractive index $n=2.4-j\Delta n$, where $j$ is the order of the step counted from the central region and $\Delta n$ is a single step-profile refractive index change. All steps have the same refractive index change, gradually changing from the un-damaged central part, $n=2.4$. The results are summarized in Table 1. These results confirm previous findings that having a gradual perturbation with smaller steps significantly increases the quality factor [17,18]. The largest $Q$ in Table 1 is 30 times better than what was achieved with a single step-profile. In future work we will investigate the optimal gradual profile in more detail.

## 5. Conclusions

We have shown that ultrahigh-$Q$ diamond-based PCS cavities can be designed. The heterostructure type cavity is formed by reducing the refractive index in the diamond slab surrounding the cavity region. Refractive index changes can be realised by low doses of damage induced via ion beam bombardment, which has the effect of lowering the refractive index. By lowering the surrounding refractive index, the cavity zone remains free of damage, which will aid the quantum properties of implanted centres. Having a single step-index profile it is possible to design cavities with $Q \sim 10^6$. Employing a slightly more complicated gradual-index profile it is possible to design cavities with $Q$-factors as high as $Q = 3 \times 10^7$. This value is around 30 times larger than previously reported designs in diamond. At the same time, the cavity volumes are still relatively small, at less than two cubic wavelengths. This unique combination is a significant enabler for diamond quantum information devices. We have also shown that this type of the cavity is extremely flexible: the range of parameters, cavity length and the refractive index change, that enable a high-$Q$ cavity is quite broad. Furthermore this type of the cavity is post-processed, enabling cavity formation only around successfully placed NV centres. Even though the radiation-induced damage can increase the losses and in general strongly affect high-$Q$ cavities, in this design this is ameliorated as most of the modal field is concentrated in the central, non-damaged, region. This combination of factors make our new diamond cavity designs ideal for applications to quantum information processing tasks such as photon sources [38,39], and gates for photonic [40] and matter-based [41] entanglement.

**Acknowledgment**

The authors thank R. Kalish and M. Draganski for helpful discussions. This work was produced with the assistance of the Australian Research Council (ARC) under the ARC Centres of Excellence Program and by an award under the Merit Allocation Scheme on the National Facility of the Australian Partnership for Advanced Computing. CUDOS (the Centre for Ultrahigh-bandwidth Devices for Optical Systems) is an ARC Centre of Excellence. A.D.G. is the recipient of an Australian Research Council Queen Elizabeth II Fellowship (project number DP0880466). A. D. G. and S. P. are supported by the Australian Research Council, the Australian Government, International Science Linkages program the Department of Education Science and Technology under the International Science Linkages scheme, and the EU 6th Framework under the EQUIND collaboration.